\documentclass[final]{ustcstep}

\usepackage[dvips]{graphicx}
\usepackage[square]{natbib}

\def\gsim{\;\lower4pt\hbox{${\buildrel\displaystyle >\over\sim}$}\;}
\def\lsim{\;\lower4pt\hbox{${\buildrel\displaystyle <\over\sim}$}\;}
\def\grls{\;\lower4pt\hbox{${\buildrel\displaystyle >\over <}$}\;}

\newcommand\addr[2]{{\footnotesize \it $^{#1}$#2}\\}

\begin{document}

\title{Waiting Times of Quasi-homologous Coronal Mass Ejections from Super Active Regions}

\author{Yuming Wang$^*$, Lijuan Liu, Chenglong Shen, Rui Liu, Pinzhong Ye, and S. Wang\\[1pt]
\addr{ }{CAS Key Laboratory of Geospace Environment, Department of Geophysics and Planetary
Sciences, University of Science and}
\addr{ }{Technology of China, Hefei, Anhui 230026, China}
\addr{*}{To whom correspondence should be addressed. E-mail: ymwang@ustc.edu.cn}}

\maketitle
\tableofcontents

\begin{abstract}
Why and how may some active regions (ARs) frequently produce coronal
mass ejections (CMEs)? It is one of the key questions to deepen our
understanding of the mechanisms and processes of energy accumulation
and sudden release in ARs and to improve our capability of space
weather prediction. Although some case studies have been made, the
question is still far from fully answered. This issue is now being
tried to address statistically through an investigation of waiting
times of quasi-homologous CMEs from super ARs in solar cycle 23. It
is found that the waiting times of quasi-homologous CMEs have a
two-component distribution with a separation at about 18 hours.
The first component is a Gaussian-like distribution with a peak
at about 7 hours, which indicates a tight physical connection between
these quasi-homologous CMEs. The likelihood of occurrences of two or more CMEs faster than 1200
km s$^{-1}$ from the same AR within 18 hours is about 20\%. Furthermore, the
correlation analysis among CME waiting times, CME speeds and CME
occurrence rates reveals that these quantities are independent to
each other, suggesting that the
perturbation by preceding CMEs rather than free energy input be the direct cause of
quasi-homologous CMEs. The peak waiting time of 7 hours probably
characterize the time scale of the growth of instabilities triggered
by preceding CMEs. This study uncovers more clues from a statistical
perspective for us to
understand quasi-homologous CMEs as well as CME-rich ARs.
\end{abstract}

\section{Introduction}
Magnetic free energy is thought to be the energy source of coronal
mass ejections (CMEs). Active regions (ARs) carry a huge amount of
free energy and therefore are the most probable place where CMEs
come out. Lots of efforts have been devoted to the triggering
mechanisms of CMEs. Flux emergence, shear motion and mass loss all
could be the initial cause of an isolated CME
\citep[e.g.,][]{Forbes_Priest_1995, Amari_etal_1996,
Chen_Shibata_2000, Manchester_2003}. No matter which one takes
effect, the determinative factor of the CME's launch is the force
balance between the inner core field and the outer overlying arcades
\citep[e.g.,][]{Wang_Zhang_2007, Liu_2007, Schrijver_2009}. Free
energy stored in the source region will be consumed when a CME
launches \citep[e.g.,][]{Sun_etal_2012}.

The picture of isolated CMEs is somewhat clear. However, it is still
a question how CMEs could lift successively in a limited region
within a relatively short interval. Usually the energy accumulation
is a gradual process in time scale of hours to days
\citep[e.g.,][]{LaBonte_etal_2007, Li_etal_2010}, while a CME is a
sudden process releasing accumulated energy in minutes. Why and how
could some ARs frequently produce CMEs? Does the occurrences of
successive CMEs from the same AR mean that the source AR accumulate
free energy quickly? The waiting time distribution of
quasi-homologous CMEs contains clues.

Homologous CMEs were defined by \citet{Zhang_Wang_2002} after
the definition of homologous flares \citep{Woodgate_etal_1984}.
Strictly speaking, homologous CMEs must originate from the same
region, have similar morphology, and be associated with homologous
flares and EUV dimmings. Here, we use the
term `quasi-homologous' to refer to successive CMEs originating from
the same ARs within a short interval, but may have different
morphology and associates.

A previous study on 15 CME-rich ARs during the ascending phase of
the last solar cycle from 1998 to 1999 have suggested that
quasi-homologous CMEs occurred at a pace of about 8 hours, and there
was at most one fast CME within 15 hours \citep{Chen_etal_2011}.
These results are important for space weather prediction, and did
imply that the accumulation rate of free energy in an AR may not
support such frequently occurrences of quasi-homologous CMEs, and
the triggering mechanisms of the first and the following CMEs are
probably different. Three scenarios were proposed to interpret the
averagely 8-hour waiting time of quasi-homologous CMEs.

Before deepening our understanding of such a phenomenon, we need to
check if a similar waiting time distribution of quasi-homologous
CMEs could be obtained for the whole solar cycle. In this paper, we
extend the period of interest to the whole solar cycle 23 from 1996
to 2006. Instead of searching all ARs and the associated CMEs, which
are too many to be identified manually, we investigate super ARs
that were reported in literatures. Super ARs are those with larger
area, stronger magnetic field and more complex pattern, and thought
to be the representative of CME producers. In the following section,
we present the selected data and the method. In
Sec.\ref{sec_results}, an analysis of waiting times of
quasi-homologous CMEs from these super ARs during the last solar
cycle is performed. Finally, conclusions and discussion is given in
the last section.

\section{Data Preparation}
\subsection{Super ARs and Associated CMEs}
Super ARs were studied by many researchers \citep{Bai_1987,
Bai_1988, Tian_etal_2002, Romano_Zuccarello_2007, Chen_etal_2011a}.
It was first defined by \citet{Bai_1987, Bai_1988} as a region
producing four and more major flares. In most studies, super ARs
were selected based on several parameters, such as the largest area
of sunspot group, the soft X-ray flare index, the 10.7 cm radio peak
flux, the short-term total solar irradiance decrease, the peak
energetic proton flux, the geomagnetic Ap index, etc. No matter
which one or more criteria are used, most selected super ARs are
CME-productive (that could be seen at the last paragraph of this
sub-section).

In our study, we focus on super ARs during solar cycle 23. Instead
of identifying super ARs by ourselves, we simple use existent lists
of super ARs in literatures. To our knowledge, there are three lists
regarding to super ARs in solar cycle 23. The first one is given by
\citet{Tian_etal_2002}, who found 16 super ARs from 1997 to 2001 base
on their selection criteria. The second one is given by
\citet{Romano_Zuccarello_2007}, which contains 26 super ARs from 2000 to 2006. The last one is in paper by
\citet{Chen_etal_2011a}, in which 12 super ARs were identified during
the last solar cycle. Since \citet{Chen_etal_2011a} used stricter
criteria, the last list is actually a subset of the other two.
Totally, we have 37 super ARs from 1996 to 2006.

\begin{figure}[tbh]
  \centering
  \includegraphics[width=\hsize]{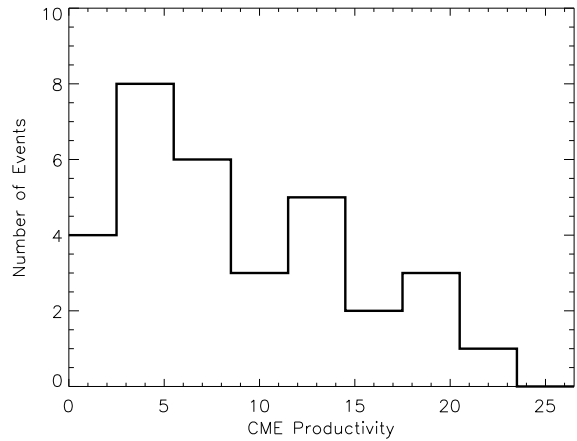}
  \caption{Distribution of CME productivities of super ARs.}\label{fg_productivity}
\end{figure}

To identify the CMEs originating from these super ARs, we examine
imaging data from Large Angle and Spectrometric Coronagraph (LASCO, \citealt{Brueckner_etal_1995})
and Extreme Ultraviolet Imaging Telescope (EIT, \citealt{Delaboudiniere_etal_1995}) onbard Solar and Heliospheric
Observatory (SOHO). The identification process is the same as that
applied by \citet{Wang_etal_2011} and \citet{Chen_etal_2011}. The CMEs
listed in the CDAW LASCO CME catalog (refer to http://cdaw.gsfc.nasa.gov/CME\_list/,
\citealt{Yashiro_etal_2004}) are our candidates. Through a careful identification,
it is found that a total of 285 CMEs are associated with these super
ARs. Figure~\ref{fg_productivity} shows the distribution of the CME productivity
of super ARs, in which the numebr of super ARs almost linearly decreases
with increasing CME number though there is a sharp decrease below the CME productivity of 3.

It should be mentioned that there are 7 super ARs with too many large data gaps
in LASCO and/or EIT observations, and therefore their CME productivity cannot
be obtained. Except them, there were 28 super ARs producing 3 or more CMEs (called
CME-rich ARs), among which 14 super ARs generated at least 10 CMEs. The other 2 super ARs
produced only one or two CMEs though sporadic data gaps existed. This fact suggests that
not all of super ARs are CME productive. But it is definite that super ARs are more
likely to be CME productive. \citet{Chen_etal_2011} identified 108
ARs during 1997--1998 and found that only 14\% of these ARs produced 3 or more
CMEs. This percentage is much lower than that for super ARs, which is about 93\% (28/30).
In this study we focus on the 28 CME-rich ARs, which produced 281 CMEs in total. A list of
all the CMEs associated with these CME-rich super ARs can be retrieved from
http://space.ustc.edu.cn/dreams/quasi-homologous\_cmes/.

\subsection{Waiting Times}
As long as there is no large data gap, we tentatively believe that
all CMEs originating from a super AR of interest are recognized
based on combined observations from SOHO LASCO and EIT. The waiting
time of each CME is obtained according to the times of first
appearance of the CME and its preceding one from the same super AR
in the field of view of LASCO/C2. However, data gaps exist, and some
CMEs may missed. If there was a large data gap between two CMEs from
the same super ARs, the waiting time of the second CME cannot be
obtained. Here, all data gaps less than 3 hours are ignored, because
it is almost impossible for a CME to stealthily escape the field of
view of LASCO in 3 hours.

Before analyze the waiting times of these CMEs from the super ARs, it has to be
noted that there are probably about 32\% of frontside CMEs missed by SOHO
\citep{Ma_etal_2010, Wang_etal_2011}. Of course, these missed CMEs might be generally
weak and faint. The statistical study by \citet{Chen_etal_2011}
have suggested that the properties of ARs have effects on the CME productivity, but
do little with the kinetic properties of CMEs. Thus, it is possible that some CMEs
originating from the super ARs are missed in our study, though such CMEs might be very
weak and erupt in a gradual manner. So far. it is hard to evaluate how significant
an influence this error will cause, and one may bear it in mind that the following
analysis is performed with a bias of normal to strong CMEs.

\section{Results}\label{sec_results}
\subsection{Waiting Time Distribution}
\begin{figure}[tbh]
  \centering
  \includegraphics[width=\hsize]{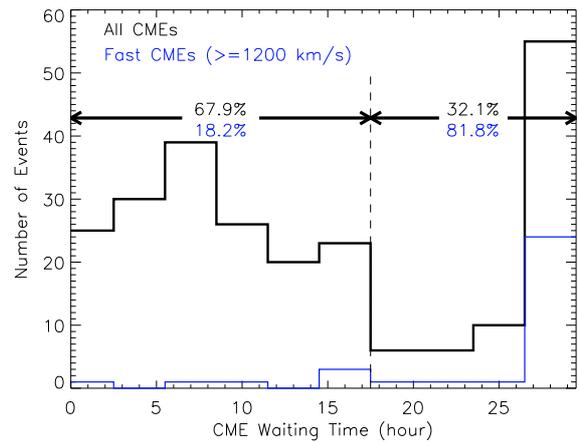}
  \caption{Waiting time distributions for all quasi-homologous CMEs (black line) and for quasi-homologous CMEs faster than 1200 km s$^{-1}$ (blue line).}\label{fg_dist}
\end{figure}

The average value of the waiting times is about 17.8 hours. The
waiting time distribution is shown in Figure~\ref{fg_dist}. Similar
to that shown in Figure 10 of the paper by \citet{Chen_etal_2011},
the distribution consists of two components. One component locates
less than 18 hours and looks like a Gaussian distribution, and the
other beyond 18 hours. For the first component distribution, the
peak waiting time is about 7 hours. In \citet{Chen_etal_2011}, the
separation of the two components of the distribution is near 15
hours, and the first component distribution peaked near 8 hours,
which are both slightly different than those obtained here. These
slight differences might be caused by the solar cycle variation.

An interesting result in \citet{Chen_etal_2011} is that any AR
cannot produce two or more CMEs faster than 800 km s$^{-1}$ within
15 hours. In other words, the time intervals between fast CMEs are
longer than 15 hours. If this result obtained during the last solar
minimum also holds for the whole solar cycle, we could expect that
any AR cannot produce two or more CMEs faster than a certain speed
threshold within 18 hours. However, such a speed threshold cannot be
found. The blue line in Figure~\ref{fg_dist} shows the waiting time
distribution for CMEs faster than 1200 km s$^{-1}$. Note that all
the slower CMEs are ignored when we calculate waiting times for CMEs
faster a certain speed threshold. Some fast CMEs did occur in the
same ARs within 18 hours. For example, there were four CMEs from the
super AR 10720 on 2005 January 15 at 06:30 UT, 23:06 UT, on January
17 at 09:30 UT and 09:54 UT, respectively, which were all faster
than 2000 km s$^{-1}$. The first two CMEs were separated by about
16.6 hours, and the other two by about only 24 minutes. These fast
CMEs caused ground-level enhancement (GLE) event
\citep[e.g.,][]{Grechnev_etal_2008}.

\begin{figure}[tbh]
  \centering
  \includegraphics[width=\hsize]{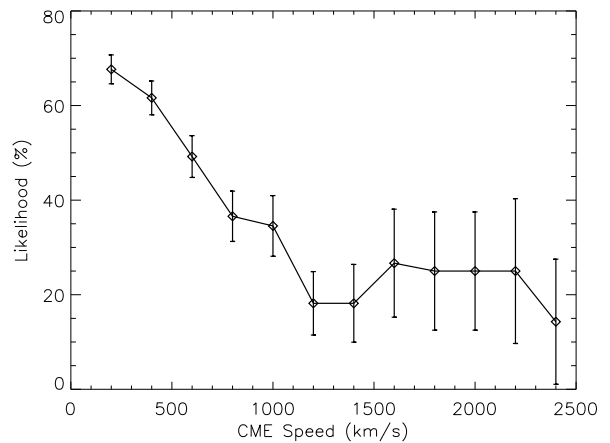}
  \caption{Dependence of likelihood of quasi-homologous CMEs occurring within 18 hours on CME speed.}\label{fg_prob}
\end{figure}

Although a similar result cannot be obtained, we find that the likelihood for
an AR producing two or more fast CMEs within 18 hours is much smaller than normal.
For all CMEs, 68\% of the waiting times are shorter than 18 hours, while for CMEs
faster than 1200 km s$^{-1}$, the fraction decreases to only about 18\%. The dependence
of the likelihood on the CME speed threshold is given in Figure~\ref{fg_prob}.
Generally, the likelihood monotonically decreases as the speed threshold increases.
When the threshold reaches to about 1200 km s$^{-1}$, the likelihood stops decreasing
and stays between 15\%--25\%, suggesting a limit likelihood of approximate 1/5.

The waiting time distribution for all CMEs from 1999 February to 2001 December
was investigated by \citet{Moon_etal_2003}, which is significantly different
from the distribution for quasi-homologous CMEs obtained here (see Figure 1 in
their paper). This difference reveals that the occurrence of CMEs follows
a Poisson process \citep{Scargle_1998, Wheatland_2000}, but that of quasi-homologous
CMEs does not. In a statistical view, we may conclude that there are tight physical
connections between quasi-homologous CMEs, but for CMEs from different source regions,
the connection is quite loose.

\subsection{Role of Free Energy Input in Causing Quasi-Homologous CMEs}
Sufficient free energy is a necessary condition for an AR to produce CMEs.
Generally, the accumulation rate of free energy could be approximately represented by
the magnetic helicity injection rate, which is another important parameter in
evaluating the productivity of ARs. Magnetic helicity measures the twists, kinks and inter-linkages
of magnetic field lines, which indicate the complexity and non-potentiality of a magnetic system.
The close relationship between the free energy and magnetic helicity could be seen from
their formulae \citep{Kusano_etal_2002}. Thus it is not surprising that a higher
injection rate of magnetic helicity often implies a higher probability of an eruptive activity, as
suggested by many studies \citep[e.g.,][]{Zhang_etal_2006a, LaBonte_etal_2007}.

However, it is still questionable if free energy input is a direct
cause of quasi-homologous CMEs. Some studies did show that CMEs
do not always occur during a quick injection of magnetic helicity or
free energy, even if the stored free energy in an AR was much higher
than that required for a CME \citep[e.g.,][]{Demoulin_Pariat_2009,
Vemareddy_etal_2012}. This issue is addressed here in a statistical
perspective from two aspects. First, we investigate the correlation
between the CME speeds and waiting times. If free energy input is a
direct cause, it is expected that there is some regulation between
CMEs' speeds and their waiting times, as a long waiting time may
lead to more free energy in an AR. This expectation is
established under the assumption that the injection rate of free
energy or magnetic helicity varies in a relatively small range for
different ARs, This assumption is statistically true based on
previous studies. For example, the statistical study by
\citet{Park_etal_2010} suggested that the magnetic helicity fluxes
in 378 ARs observed by SOHO/MDI were on the order of about $10^{40}$
Mx h$^{-1}$, especially for those ARs with large magnetic flux (see,
e.g., Fig.1, 3 and 4 in their paper). The value does not change much
even if deriving from higher-resolution data from SDO/HMI, e.g., the
helicity injection rate in AR 11158 and 11166
\citep{Vemareddy_etal_2012}.

\begin{figure}[tbh]
  \centering
  \includegraphics[width=\hsize]{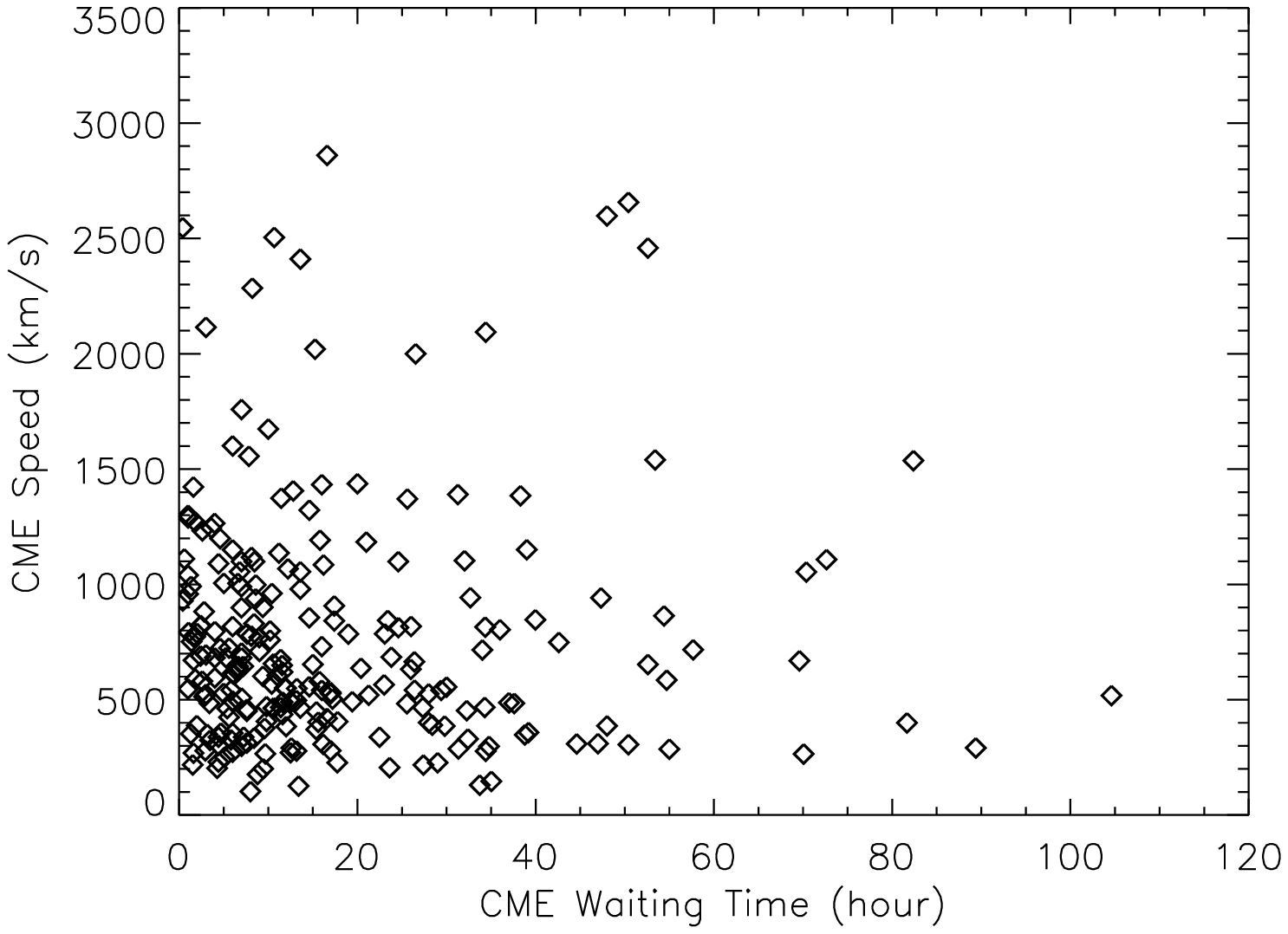}\\
  \includegraphics[width=\hsize]{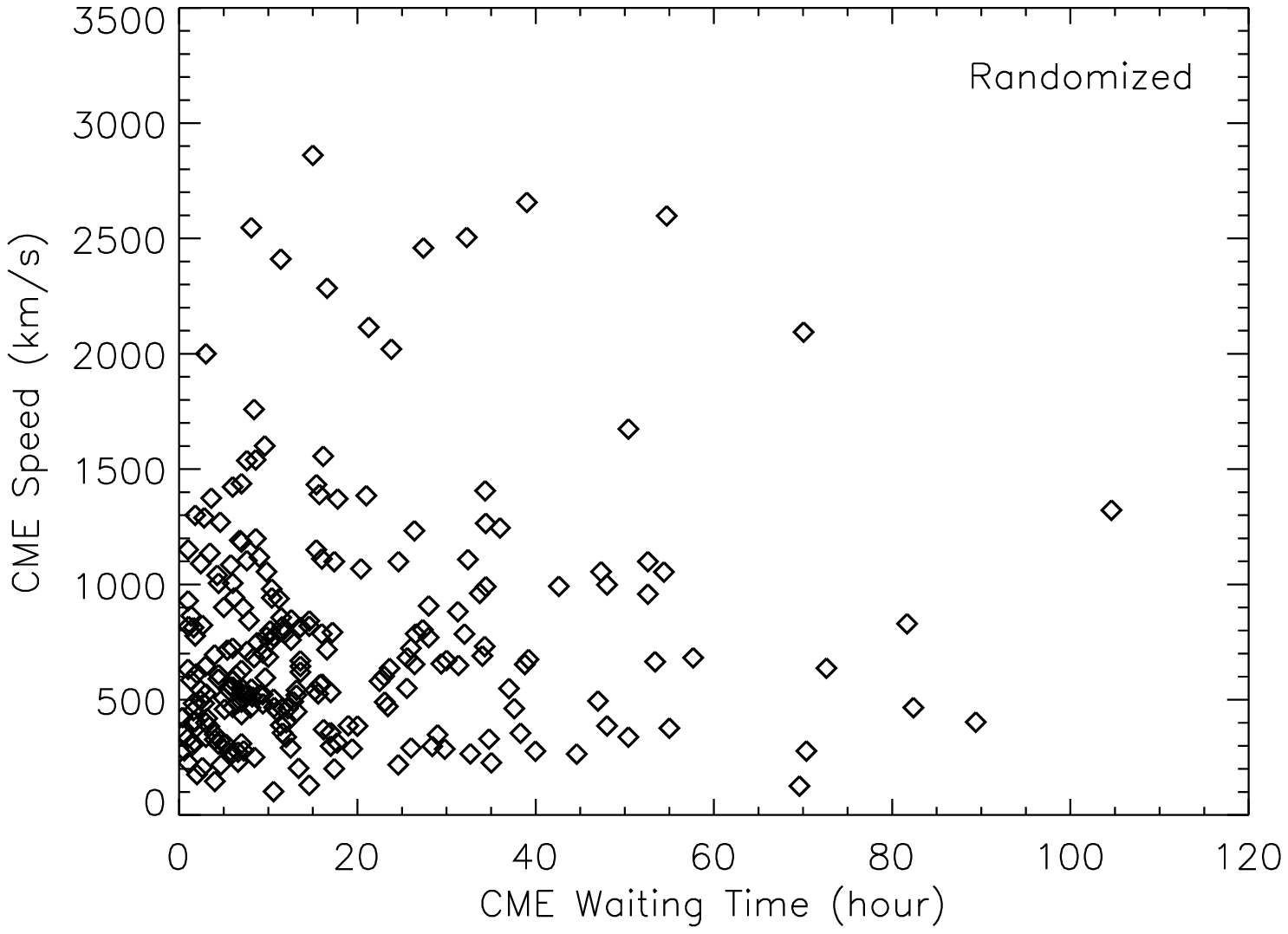}
  \caption{{\it Upper Panel}: Scattering plot of CME speeds versus CME waiting times. {\it Lower Panel}:
Same as {\it Upper Panel}, but the association between them is randomized.}\label{fg_wtdep1}
\end{figure}

Figure~\ref{fg_wtdep1}a shows the dependence of CME speed on the
waiting time. Overall, no clear correlation could be found between them, except that
there is seemingly an upper limit in CME speed depending on the CME waiting time.
However, although the distribution is statistically true, it does
not imply that an AR is difficult to produce a fast CME if it had waited too long.
It is a result simply from a combination of two Gaussian-like distributions.
The CME waiting time is a Gaussian-like distribution, at least for the first component
(as shown in Figure~\ref{fg_dist}). The CME speed is actually also a Gaussian-like distribution.
If the two quantities are independent, the 2-D distribution composed by them is
like that shown in Figure~\ref{fg_wtdep1}a. As a test, Figure~\ref{fg_wtdep1}b shows the
distribution, in which the CME speeds in our sample are randomly associated with the
CME waiting times. The two distributions given in Figure~\ref{fg_wtdep1}a and \ref{fg_wtdep1}b
are quite similar. It reflects that the CME speed is independent on the CME waiting time.

\begin{figure}[tbh]
  \centering
  \includegraphics[width=\hsize]{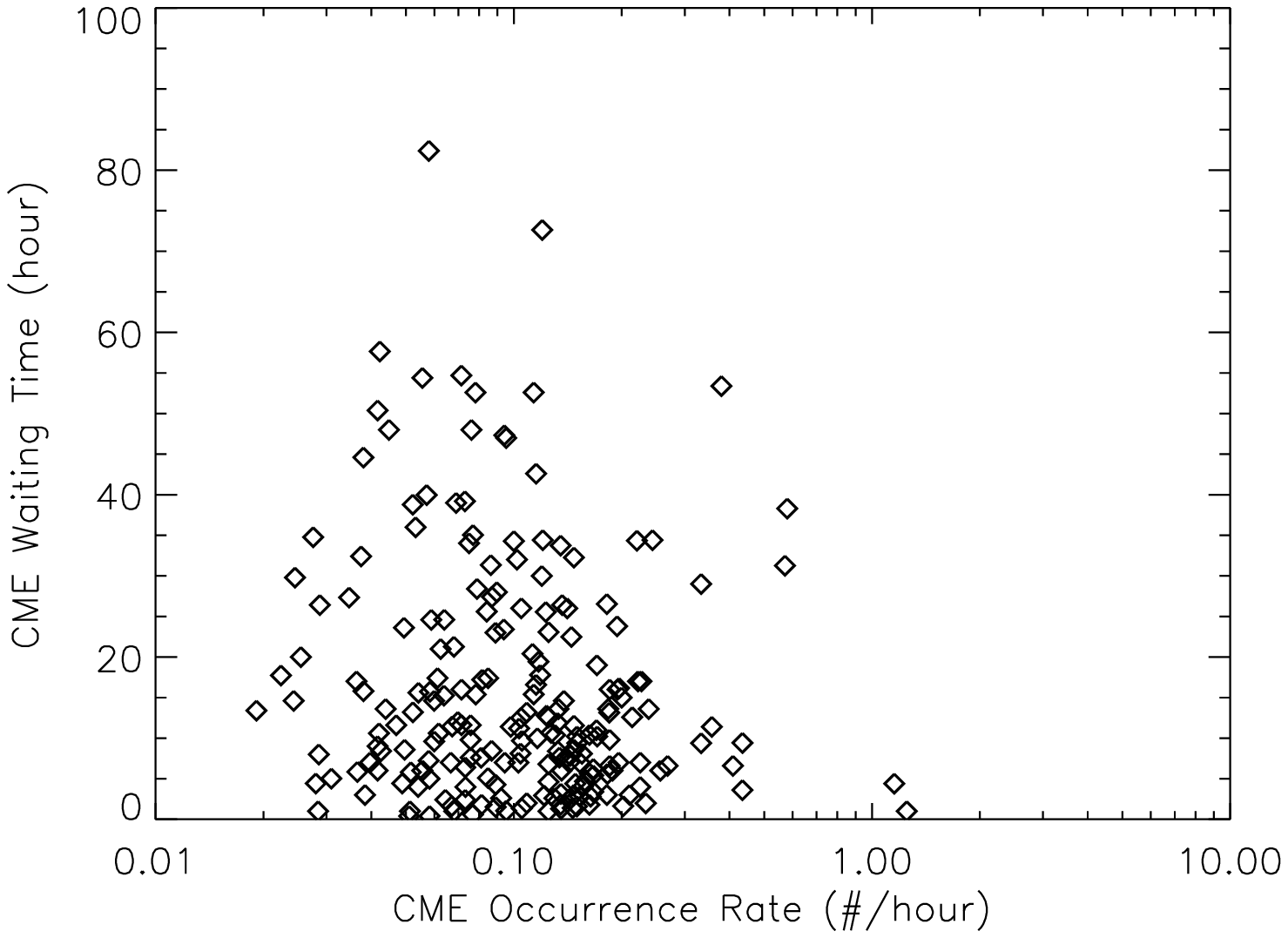}\\
  \includegraphics[width=\hsize]{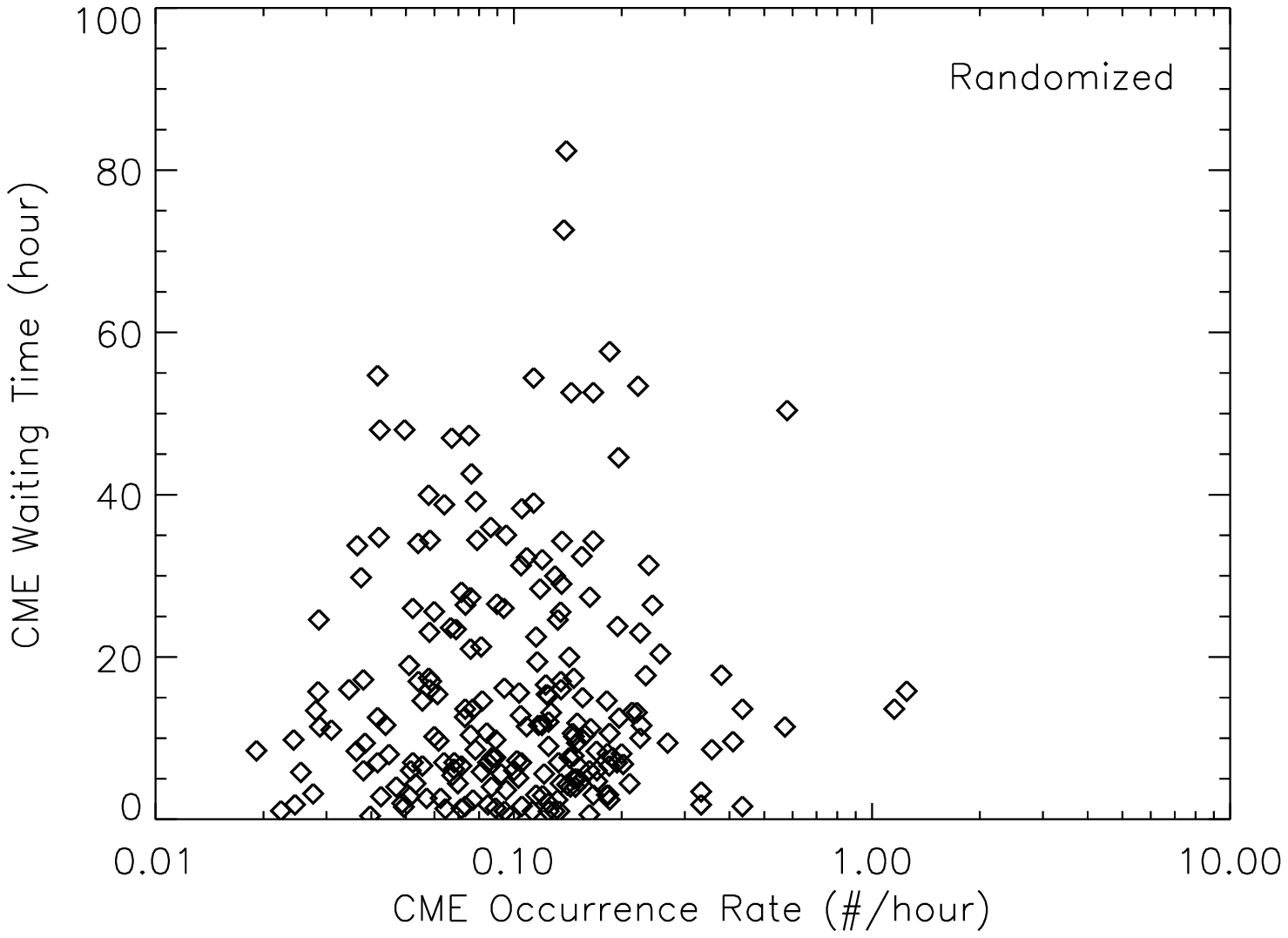}
  \caption{{\it Upper Panel}: Scattering plot of CME waiting times versus CME occurrence rate in the past 18 hours.
{\it Lower Panel}: Same as {\it Upper Panel}, but the association between them is randomized.}\label{fg_wtdep2}
\end{figure}

Second, we check if the waiting time of a CME depends on the CME
occurrence rate in the past 18 hours before its preceding CME.
Figure~\ref{fg_wtdep2}a shows the scattering plot between them.
Apparently, a low or high CME occurrence rate may lead to a short
waiting time of the next CME, and a long waiting time tends to
appear when the CME occurrence rate is around 0.1 per hour. However,
similar to the previous one, this distribution is also just a
manifestation of probability, and contains less physical meaning. If
we randomly associate the CME waiting times with the occurrence
rates, a possible distribution of the data points is like that shown
in Figure~\ref{fg_wtdep2}b, which is statistically same as that in
Figure~\ref{fg_wtdep2}a. Thus the CME waiting time is independent on
the previous CME occurrence rate. Both results suggest that free
energy input is not a direct cause of quasi-homologous CMEs though
sufficient free energy is a necessary condition for an AR to produce
CMEs.

\section{Summary and discussion}
In summary, by investigating 281 quasi-homologous CMEs originating
from 28 CME-rich super ARs over the last solar cycle, we find a
two-component distribution of their waiting times with the
separation of the two components at about 18 hours and the peak
waiting time of the first component at about 7 hours. These results
suggest a close physical connection between quasi-homologous CMEs
which fall in the first component. Furthermore, the likelihood of
occurrences of two or more fast CMEs within 18 hours decreases as
CME speed increases. A limit likelihood of about 20\% is reached
when CME speed is larger than 1200 km s$^{-1}$.

The correlation analysis among CME waiting times, CME speeds and
previous CME occurrence rates shows us the statistical evidence that
the free energy input is
not a direct cause of quasi-homologous CMEs. It is well known that
that the free energy stored in ARs may be much higher than that could
be consumed by one single CME \citep[e.g.,][]{Sun_etal_2012}. Thus the
direct cause of quasi-homologous CMEs is not the quick
re-fill of free energy after preceding CMEs, but the perturbation by
preceding CMEs, which may lower the threshold of eruption or
trigger instabilities to cause the next CME. Pre-eruption flux rope
is precisely balanced by outward force
from inner core field and inward force from overlying arcades
\citep[e.g.,][]{Torok_Kliem_2005, Wang_Zhang_2007, Liu_2007}. A CME
may reduce the constraint of its nearby flux rope system by removing
overlying arcades, and cause the balance broken. As shown
in the numerical simulation by \citet{Torok_etal_2011}, which was
designed to study the physical mechanism of a global sympathetic
eruptions on 2010 August 1 \citep{Schrijver_Title_2011}, the second and
third eruptions were actually caused by preceding eruptions. In their
eruption processes, the preceding eruption caused the overlying arcades
reduced through reconnection, and then instability developed. A similar
result was obtained in the simulation by \citet{Bemporad_etal_2012}, in
which the second CME was caused by the rearrangement of coronal magnetic
field after the first CME.

Connecting the above picture to the peak waiting time of 7 hours, we
may speculate that the 7-hour waiting time probably characterizes the
average time scale of the growth of instabilities.
In our previous work \citep{Chen_etal_2011}, we
proposed three scenarios to interpret the peak waiting time. Here we
may tentatively narrow down them to the last two, in which
quasi-homologous CMEs probably hatched from a long magnetic flux
system or different magnetic flux systems in one AR. A simple/small
AR should be difficult to frequently produce CMEs. A detailed
investigation on this point is worthy to be carried out in future
work.

\acknowledgments{ We acknowledge the use of the data from SOHO
LASCO, EIT and MDI and the CDAW CME catalog. SOHO is a project of
international cooperation between ESA and NASA. This work is
supported by grants from CAS (Key Research Program KZZD-EW-01 and
100-Talent Program), NSFC (41131065, 40904046, 40874075, 41121003,
41274173 and 41222031), 973 key project (2011CB811403), MOEC
(20113402110001) and the fundamental research funds for the central
universities. RL is also supported by NSF (AGS-1153226).}

\bibliographystyle{agufull}
\bibliography{../../ahareference}

\end{document}